# Invariant Stream Generators using Automatic Abstract Transformers based on a Decidable Logic[*]

Pierre-Loïc Garoche[1,2], Temesghen Kahsai[2] and Cesare Tinelli[2]

[1] Onera, the French Aerospace Lab, France
[2] The University of Iowa

**Abstract.** The use of formal analysis tools on models or source code often requires the availability of auxiliary invariants about the studied system. Abstract interpretation is currently one of the best approaches to discover useful invariants, especially numerical ones. However, its application is limited by two orthogonal issues: (i) developing an abstract interpretation is often non-trivial; each transfer function of the system has to be represented at the abstract level, depending on the abstract domain used; (ii) with precise but costly abstract domains, the information computed by the abstract interpreter can be used only once a post fix point has been reached; something that may take a long time for very large system analysis or with delayed widening to improve precision. This paper proposes a new, completely automatic, method to build abstract interpreters. One of its nice features is that its produced interpreters can provide sound invariants of the analyzed system before reaching the end of the post fix point computation, and so act as on-the-fly invariant generators.

## 1    Introduction and Motivation

Theoretical frameworks such as abstract interpretation and symbolic (specifically, logic-based) model checking have led in the last few years to the development of analysis tools that are starting to have a strong practical impact on the development of real word software, in particular for safety- or mission-critical systems. Interestingly, current abstract interpretation and model checking techniques exhibit complementary strengths and weaknesses. Model checking techniques so far have been stronger on software that is mostly control-driven and not heavily data-dependent. To be effective with data-dependent programs, these techniques may require programs to be judiciously annotated with data invariants. Also, model checking has been traditionally limited to finite-state systems, although new approaches relying on solvers for Satisfiability Modulo Theories (SMT) are starting to remove that limitation.

Dually, abstract interpretation techniques are quite effective on data-dependent programs, in particular numerical ones, requiring in principle no program annotations. On the other hand, they have more difficulties in dealing with control aspects. Also, although abstract interpretation is a very general framework, most of its applications focus on the analysis of source code. Even tools, such as Nbac [13], that target software artifacts at a higher level of abstraction (e.g., software models expressed in dataflow specification languages) do not analyze those artifacts directly and work instead with their compilation into C code. This is possibly a consequence of the fact that developing an abstract interpreter for a complete language can be time consuming: even if a large set of abstract domains, such as those provided by the APRON library [14], is readily available, it requires substantial work to define sound abstract transformers for every construct of the target language. Another limitation of current abstract interpretation techniques is that they typically rely on Kleene-style fix point algorithms to construct an abstract semantics of the program under analysis. The properties of such semantics, characterized by the concretization of a post fix point of an abstract trasformer, can be obtained only once the post fix point has been (completely) computed. Depending on the widening strategies used or, in general, the complexity of the abstractions and the semantics considered, one may have to wait a long time to get any information at all from the analysis of the program.

[*]This work was partially supported by AFOSR grant #AF9550-09-1-0517, and by the FN-RAE Cavale project.

*Contributions.*  In this work we try to address some of the issues above by combining techniques from abstract interpretation and logic-based model checking. Specifically, we propose a general method for the automatic definition of abstract interpreters that compute numerical invariants of transition systems. We rely on the possibility of encoding the transition system in a decidable logic—such as those typically used by SMT-based model checkers—to compute transformers for an abstract interpreter *completely automatically*. Our method has the significant added benefit that the abstract interpreter can be instrumented to generate system invariants on the fly, during its iterative computation of a post fix point. A prototype implementation of the method provides initial evidence of the feasibility of our approach and the usefulness of its incremental invariant generation feature.

*Significance.*  While motivated by practical issues (namely, the generation of auxiliary invariants for a *k*-induction model checker) the current work is more general and can be adapted to a wide variety of contexts. It only requires that the transition system semantics be expressible in a decidable logics with an efficient solver, such as SAT or SMT solvers, and that the elements of the chosen abstract domain be effectively representable in that logic (as discussed later in more detail). Such requirements are satisfied by a large number of abstract domains used in current practice. As a consequence, we believe that our approach could help considerably in expanding the reach of abstract interpretation techniques to a variety of target languages, as well as facilitate their integration with complementary techniques such as model checking ones.

*Related work.*  With the current efficiency of SMT solvers on the one hand and the ability of abstract interpretation to compute numerical invariants on the other, the issue of combining SMT and AI is receiving increased attention. In [7], Cousot, Cousot and Mauborgne draw a parallel between SMT-based reasoning and abstract interpretation. They identify the Nelson-Oppen procedure as a reduced product over different interpretations. While this work is more general, it allows one to understand ours as follow: the concrete domain presented in Figure 2 is an abstract logical domain, our concrete

transformer—computed with the aid of an SMT solver—can be understood as an over-approximation of the concrete transition relation in this abstract logical domain. The abstraction we built amounts to compute a reduction between a logical and an algebraic domain, as suggested in [7, §6]. Comparable work in [22], gives an overview of techniques embedding logical predicates as elements of *logical lattices*. Some SMT theories are then formalized within this abstract interpretation view of the analysis: uninterpreted function symbols, linear arithmetic, and their combination.

Another more practical approach, by Monniaux and Gonnord [20], uses bounded reachability with an SMT solver to compute a chaotic iteration strategy. The solver identifies the equation that needs propagating in order to achieve a better widening. However, unlike ours, this solution does not rely on the actual (counter-)models synthesized by the SMT solver. In [10], an SMT solver is used to choose among different strategies in an iteration-based policy analysis. The solver identifies the next strategy that will improve the current abstract property. The latter two works rely on SMT solvers to help the fix point computation but do not rely on the SMT-based concrete semantics to compute the abstract property.

Another line of work addresses the embedding of abstract interpretation into logical frameworks. In [11], the authors proposes an abstract domain with quantification over a specific pattern of properties. They provide generic transfer functions and lattice operators that enable the representation of properties like $\bigwedge (P_i \Rightarrow Q_i)$.

Also related is Monniaux's automatic modular abstraction for linear constraints [19]. A predicate transformer is defined using quantifier elimination over the semantics of C statements, as in an axiomatic semantics (weakest precondition or strongest postcondition). The transformer is exact for the linear template abstractions considered. It is however not clear how this approach can scale to a complete program analysis, since the use of quantifier elimination on a complete transition system is not usually feasible. In [19] the analyzed blocks are small functions used in a symbol library for Lustre/S-cade.

## 2    Formal Preliminaries

We rely on basic notions and results from abstract interpretation [3,4,5, e.g.]. We introduce below those that are most relevant to this work, to have a more self-contained presentation. Similarly, we also introduce relevant notions from symbolic logic and automated reasoning.

As customary, we will model computational systems as transition systems. A *transition system $S$* is a triple $(\mathbf{Q}, \mathbf{I}, \rightsquigarrow)$ where $\mathbf{Q}$ is a set of *states*, the *state space*; $\mathbf{I} \subseteq \mathbf{Q}$ is the set of $S$'s *initial states*; and $\rightsquigarrow \subseteq \mathbf{Q} \times \mathbf{Q}$ is $S$'s *transition relation*. A state $q \in \mathbf{Q}$ is *reachable* if $q \in \mathbf{I}$ or $q' \rightsquigarrow q$ for some reachable state $q'$.

**Abstract Interpretation**  Abstract interpretation allows one to analyze a transition system $S = (\mathbf{Q}, \mathbf{I}, \rightsquigarrow)$ by first defining a *concrete domain* for $S$, a partially ordered set $\langle D, \sqsubseteq \rangle$, and a *concrete transformer*, a monotonic function $f : D \rightarrow D$. In this paper we will focus on the *collecting semantics*

$$\mathbb{S} \stackrel{\text{def}}{=} \mathrm{lfp}_{\mathbf{I}}^{\sqsubseteq}(f)$$

of $S$ where $D = \wp(\mathbf{Q})$, $\subseteq$ is set inclusion, $f(X) = X \cup \{x' \mid x \in X, \, x \rightsquigarrow x'\}$ and $\text{lfp}_{\mathbf{I}}^{\subseteq}(f)$ is the least-fix point of $f$ greater than $\mathbf{I}$, obtained as the stationary limit of the ascending sequence $X_0 \subseteq X_1 \subseteq \ldots$ with $X_0 = \mathbf{I}$ and $X_n = f(X_{n-1})$ for all $n > 0$. However, our work could be extended to other semantics, such as trace semantics.

The second step in abstract interpretation consists in providing an abstract representation of the chosen concrete domain, the *abstract domain*, given by another partial order $\langle D^\#, \sqsubseteq^\# \rangle$ (typically a complete partial order if $\langle D, \subseteq \rangle$ is one). The two domains are related by an *abstraction function* $\alpha : D \mapsto D^\#$ and a *concretization function* $\gamma : D^\# \mapsto D$ that respectively associate an *abstract element*, a member of $D^\#$, to each *concrete element*, a member of $D$, and vice versa. We call an *abstract transformer* any monotonic function $g : D^\# \to D^\#$. A *good* abstraction function is closed under intersection, to ensure the existence of the best abstraction for each concrete element. Galois connection-based abstractions satisfy this desideratum.

**Definition 1 (Galois connection).** *Two functions $\alpha : D \to D^\#$ and $\gamma : D^\# \to D$ form a* Galois connection *between two lattices $\langle D, \subseteq \rangle$ and $\langle D^\#, \sqsubseteq \rangle$, which we denote by $\alpha : \langle D, \subseteq \rangle \leftrightarrows \langle D^\#, \sqsubseteq \rangle : \gamma$, if (i) both $\alpha$ and $\gamma$ are monotonic; (ii) for all $y \in D^\#$, $\alpha(\gamma(y)) \sqsubseteq y$; and (iii) for all $x \in D$, $x \subseteq \gamma(\alpha(x))$.*

We will rely on the following important property of Galois connections.

**Proposition 1 (Unique adjoint in a Galois connection [4]).** *If $\alpha : \langle D, \subseteq \rangle \leftrightarrows \langle D^\#, \sqsubseteq \rangle : \gamma$ then (i) for all $x \in D$, $\alpha(x) = \bigsqcap \{y \mid x \subseteq \gamma(y)\}$; (ii) for all $y \in D^\#$, $\gamma(y) = \bigcup \{x \mid \alpha(x) \sqsubseteq y\}$; where $\bigsqcap$ and $\bigcup$ denote respectively the greatest lower bound and the lowest upper bound operators in the two lattices.*

In a Galois connection, abstract transformers can be related to concrete ones according to the following notion of sound approximation.

**Definition 2.** *If $\alpha : \langle D, \subseteq \rangle \leftrightarrows \langle D^\#, \sqsubseteq \rangle : \gamma$, an abstract transformer $f^\# : D^\# \to D^\#$ is a* sound approximation *of a concrete transformer $f : D \to D$ if for all $x \in D$, $(\alpha \circ f)(x) \sqsubseteq (f^\# \circ \alpha)(x)$ or, equivalently, for all $y \in D^\#$, $(f \circ \gamma)(y) \sqsubseteq (\gamma \circ f^\#)(y)$.*

Abstract transformers in the function space $(D^\# \to D^\#)$ are partially ordered by the point-wise extension of $\sqsubseteq$, which we denote also by $\sqsubseteq$. In Galois connections, the set of sound approximations of a concrete transformer has a smallest element wrt $\sqsubseteq$.

**Proposition 2 (Best sound approximation [4]).** *If $\alpha : \langle D, \subseteq \rangle \leftrightarrows \langle D^\#, \sqsubseteq \rangle : \gamma$, an abstract transformer $f^\# : D^\# \to D^\#$ is a sound approximation, wrt this Galois connection, of a concrete transformer $f : D \to D$ iff $\alpha \circ f \circ \gamma \sqsubseteq f^\#$.*

The property above implies that $\alpha \circ f \circ \gamma$ is the *best abstract transformer* for $f$, in the sense of being its tightest sound approximation.

**First-order logic.** Our method works with several logics that can be more or less directly embedded in many-sorted first-order logic with equality [9,18] (including propositional logic and quantified Boolean logic). For generality then, we present our work in terms of that logic. We fix an infinite set $\mathbf{S}$ of *sort symbols*. For each $\sigma \in \mathbf{S}$, we also fix an infinite set $\mathbf{X}_\sigma$ of *variables (of sort $\sigma$)*, with $\mathbf{X}_{\sigma_1}$ disjoint from $\mathbf{X}_{\sigma_2}$ for all distinct $\sigma_1, \sigma_2 \in \mathbf{S}$, and let $\mathbf{X} = \bigcup_{\sigma \in \mathbf{S}} \mathbf{X}_\sigma$. A many-sorted *signature* $\Sigma$ consists of a set $\Sigma^\mathrm{S} \subseteq S$ of sort symbols, a set $\Sigma^\mathrm{P}$ of *(sorted) predicate symbols* $p^{\sigma_1 \cdots \sigma_n}$, a set $\Sigma^\mathrm{F}$ of *(sorted) function symbols*, $f^{\sigma_1 \cdots \sigma_n \sigma}$, where $n \geq 0$ and $\sigma_1, \ldots, \sigma_n, \sigma \in \Sigma^\mathrm{S}$. We drop the sort superscript from function or predicate symbols when it is clear from context or unimportant.

For each $\sigma \in \Sigma^\mathrm{S}$, a *($\Sigma$-)term of sort $\sigma$* is either a variable $x \in \mathbf{X}_\sigma$ or an expression of the form $f^{\sigma_1 \cdots \sigma_n \sigma}(t_1, \ldots, t_n)$ where $f^{\sigma_1 \cdots \sigma_n \sigma} \in \Sigma^\mathrm{F}$ and $t_i$ is a term of sort $\sigma_i$ for $i = 1, \ldots, n$. An *atomic ($\Sigma$-)formula* is an expression of the form $t_1 = t_2$ where $t_1$ and $t_2$ are terms of the same sort,[3] or one of the form $p^{\sigma_1 \cdots \sigma_n}(t_1, \ldots, t_n)$ with $n \geq 0$ where $p^{\sigma_1 \cdots \sigma_n} \in \Sigma^\mathrm{P}$ and $t_i$ is a $\Sigma$-term of sort $\sigma_i$ for $i = 1, \ldots, n$. Non-atomic formulas with the usual Boolean connectives (false, $\neg$, $\vee$, $\wedge$, $\Rightarrow$, . . .) and quantifiers ($\forall$, $\exists$) are defined as expected. Free and bound occurrences of a variable in a formula are also defined as usual. If $F$ is a $\Sigma$-formula and $(x_1, \ldots, x_n)$ a tuple of distinct variables, we write $F[x_1, \ldots, x_n]$ to express that the free variables of $F$ are in $(x_1, \ldots, x_n)$; furthermore, if $t_1, \ldots, t_n$ are terms with each $t_i$ of the same sort as $x_i$, we write $F[t_1, \ldots, t_n]$ to denote the formula obtained from $F[x_1, \ldots, x_n]$ by simultaneously replacing each occurrence of $x_i$ in $F$ by $t_i$, for $i = 1, \ldots, k$. We denote finite tuples of elements by letters in bold font, and use comma (,) for tuple concatenation.

For each signature $\Sigma$, a *$\Sigma$-interpretation $\mathcal{M}$* is a mathematical structure that maps: each $\sigma \in \Sigma^\mathrm{S}$ to a non-empty set $M_\sigma$, the *domain* of $\sigma$ in $\mathcal{M}$; each $x \in \mathbf{X}$ of sort $\sigma$ to an element $x^\mathcal{M} \in M_\sigma$; each $f^{\sigma_1 \cdots \sigma_n \sigma} \in \Sigma^\mathrm{F}$ to a total function $f^\mathcal{M} : M_{\sigma_1} \times \cdots \times M_{\sigma_n} \to M_\sigma$ (and in particular each constant $c$ of sort $\sigma$ to an element $c^\mathcal{M} \in M_\sigma$); each $p^{\sigma_1 \cdots \sigma_n} \in \Sigma^\mathrm{P}$ to a set $p^\mathcal{M} \subseteq M_{\sigma_1} \times \cdots \times M_{\sigma_n}$.

Every $\Sigma$-interpretation $\mathcal{M}$ over some $X \subseteq \mathbf{X}$ induces a unique mapping $(\_)^\mathcal{M}$ from $\Sigma$-terms $f(t_1, \ldots, t_n)$ with variables in $X$ to elements of sort domains such that $(f(t_1, \ldots, t_n))^\mathcal{M} = f^\mathcal{M}(t_1^\mathcal{M}, \ldots, t_n^\mathcal{M})$. A satisfiability relation $\models$ between such interpretations and $\Sigma$-formulas with variables in $X$ can defined inductively as usual. A $\Sigma$-interpretation $\mathcal{M}$ *satisfies* a $\Sigma$-formula $F$ if $\mathcal{M} \models F$. A $\Sigma$-formula $F$ is *satisfiable* if it is satisfied by some $\Sigma$-interpretation. A set $\Gamma$ of $\Sigma$-formulas is *satisfiable* if there is a $\Sigma$-interpretation that satisfies every formula in $\Gamma$.

We are not generally interested in arbitrary formulas and interpretations but in specific sets of $\Sigma$-formulas and specific classes of $\Sigma$-interpretations, for some signature $\Sigma$. We collect these restrictions in the notion of a (sub)logic (of many-sorted logic). More precisely, a *logic* is a triple $\mathcal{L} = (\Sigma, \mathbf{F}, \mathbf{M})$ where $\Sigma$ is a signature; $\mathbf{F}$, the *language* of $\mathcal{L}$, is a set of $\Sigma$-formulas; and $\mathbf{M}$ is a class of $\Sigma$-interpretations, the *models* of $\mathcal{L}$, that is closed under *variable reassignment*, i.e., $\mathcal{M}[x \mapsto a] \in \mathbf{M}$ for all $\mathcal{M} \in \mathbf{M}$, all variables $x$ of sort $\sigma$ and all $a \in M_\sigma$, where $\mathcal{M}[x \mapsto a]$ is the $\Sigma$-interpretation that maps $x$ to $a$ and is otherwise identical to $\mathcal{M}$. A formula $F[\boldsymbol{x}]$ of $\mathcal{L}$ is *satisfiable (resp., unsatisfiable) in $\mathcal{L}$* if it is satisfied by some (resp., no) interpretation in $\mathbf{M}$. A set $\Gamma$ of formulas *entails in $\mathcal{L}$* a $\Sigma$-formula $F$, written $\Gamma \models_\mathcal{L} F$, if $\Gamma \cup \{F\} \in \mathbf{F}$ and every interpretation in $\mathbf{M}$ that satisfies all formulas in $\Gamma$ satisfies $F$ as well. The set $\Gamma$ is *satisfiable in $\mathcal{L}$* if $\Gamma \not\models_\mathcal{L}$ false. Two formulas $F$ and $G$ are *equivalent in $\mathcal{L}$* if $F \models_\mathcal{L} G$ and $G \models_\mathcal{L} F$.[4]

---

[3] We will use = also to denote equality at the meta-level, relying on context to disambiguate.

[4] All these notions reduce to the corresponding standard ones in many-sorted logic when $\mathbf{F}$ is the set of all $\Sigma$-formulas and $\mathbf{M}$ the class of all $\Sigma$-interpretations.

## 3   Automatic definition of a computable abstract transformer via an encoding in a decidable logic

For the rest of the paper we fix a transition system $S = (\mathbf{Q}, \mathbf{I}, \rightsquigarrow)$ and its collecting semantics $\mathbb{S} = \mathrm{lfp}_\mathbf{I}^\subseteq(f)$ introduced earlier, which coincides with the set of reachable states of $S$. Our main concern will be how to define an abstract counterpart $f_\mathbf{A}$ of $f$ in a suitable Galois connection $\alpha : \langle \wp(\mathbf{Q}), \subseteq \rangle \leftrightarrows \langle \mathbf{A}, \sqsubseteq \rangle_\mathbf{A} : \gamma$ so that we can define $S$'s abstract semantics as

$$\mathbb{S}^\# \stackrel{\mathrm{def}}{=} \mathrm{lfp}_{\mathbf{I}_\mathbf{A}}^{\sqsubseteq_\mathbf{A}}(f_\mathbf{A})$$

where $\mathbf{I}_\mathbf{A}$ is in turn a suitable abstraction of $\mathbf{I}$. By well-known results by Kleene and Cousot and Cousot [3,4], the fix point above can be computed or over-approximated so that its concretization by $\gamma$ is a sound approximation of the concrete fix point $\mathbb{S}$.

A main issue when using abstract interpretation in general is to how define $f_\mathbf{A}$. In practice, when the transition system is generated, as is often the case, by a program in a certain programming language, the concrete transformer $f$ is defined constructively in terms of the language's idioms (e.g., assignment, loop and conditional statements for imperative languages) and memory model (e.g., heap, stack, etc.). The corresponding abstract transformer must then handle all those those constructs as well, and reflect their respective actions in the abstract domain $D_\mathbf{A}$.

When the abstraction is defined via the unique adjoint property of the Galois connection, the definition of $f_\mathbf{A}$ is usually a manual, laborious chore. One has to design the transformer in detail and then prove it sound, by showing that $f(X) \in \gamma(f_\mathbf{A}(a))$ for all $a \in \mathbf{A}$ and $X \in \gamma(a)$. We present a method that, under the right conditions, can instead compute a sound abstraction of $f$ completely automatically. The method is applicable when the transition system and the concrete and abstract domains can be encoded, as explained later, in a logic $\mathcal{L}$ satisfying a number of requirements. For generality, we will describe our method in terms of an arbitrary logic $\mathcal{L}$ satisfying those requirements. To have a clue, however, depending on the concrete domain, possible examples of $\mathcal{L}$ would be propositional logic or several of the many logics used in SMT: linear real arithmetic, linear integer arithmetic with arrays, and so on.

**Logic requirements.** We assume a logic $\mathcal{L} = (\Sigma, \mathbf{F}, \mathbf{M})$ with a decidable entailment relation $\models_\mathcal{L}$ and a language $\mathbf{F}$ closed under all the Boolean operators.[5] For each sort $\sigma$ in $\mathcal{L}$, we distinguish a set $V_\sigma$ of variable-free terms, which we call *values*, such that $\models_\mathcal{L} \neg(v_1 = v_2)$ for each distinct $v_1, v_2 \in V_\sigma$. Examples of values would be integer constants, selected terms of the form $n/m$ where $n$ is an integer constant and $m$ a non-zero numeral, and so on. We assume that the satisfiable formulas of $\mathcal{L}$ are satisfied by values, that is, for every formula $F[\boldsymbol{y}]$ (with free variables from $\boldsymbol{y}$) satisfiable in a model $\mathcal{M}$ of $\mathcal{L}$ there is a value tuple $\boldsymbol{v}$ such that $F[\boldsymbol{v}]$ is satisfiable in $\mathcal{M}$.

We assume a total surjective encoding of $S$'s state space $\mathbf{Q}$ to $n$-tuples of values in the sense above, for some fixed $n$ (where each $n$-tuple encodes a state). Depending on $\mathcal{L}$, states may be encoded, for instance, as tuples of Boolean constants or integer constants, or mixed tuples of Boolean, integer and rational constants, and so on. Because of this encoding, from now on *we will identify states with tuples of values*. Note that, thanks to our various assumptions, each formula $F[\mathbf{y}_1, \ldots, \mathbf{y}_k]$ in $k \cdot n$ variables denotes a subset of $\mathbf{Q}^k$, namely the set of all $k$-tuples of states that satisfy $F$. We call that set the *extension* of $F$ and define it formally as follows:

$$\llbracket F \rrbracket \overset{\text{def}}{=} \{(\mathbf{v}_1, \ldots, \mathbf{v}_k) \in \mathbf{Q}^k \mid F[\mathbf{v}_1, \ldots, \mathbf{v}_k] \text{ is satisfiable in } \mathcal{L}\} \, .$$

---

[5] The latter is mostly to simplify the exposition. Weaker assumptions are possible.

We will refer to formulas like $F$ above as *state formulas* and say they are *satisfied* by the state sequences in $\llbracket F \rrbracket$. For each state $\mathbf{v} = (v_1, \ldots, v_n) \in \mathbf{Q}$ and tuple $\mathbf{x} = (x_1, \ldots, x_n)$ of distinct variables of corresponding sort, we denote by $A_{\mathbf{v}}$ the *assignment formula* $x_1 = v_1 \wedge \cdots \wedge x_n = v_n$, which is satisfied exactly by $\mathbf{v}$.

Finally, we assume the existence of an *encoding of $S$ in $\mathcal{L}$*, a pair of formulas of $\mathcal{L}$,

$$(I[\mathbf{x}], \, T[\mathbf{x}, \mathbf{x}'])$$

with $\mathbf{x}$ and $\mathbf{x}'$ both of size $n$, where $I[\mathbf{x}]$ is a formula satisfied exactly by the initial states of $S$, and $T[\mathbf{x}, \mathbf{x}']$ is a formula satisfied by two reachable states $\mathbf{v}, \mathbf{v}'$ iff $\mathbf{v} \rightsquigarrow \mathbf{v}'$.

**First abstraction: from sets of states to formulas of $\mathcal{L}$.** We start with an intermediate abstraction that maps sets of states to formulas representing those states. To do that, we extend the language of $\mathcal{L}$ by closing it under a disjunction operator $\bigvee$ that applies to (possibly infinite) sets of formulas of $\mathcal{L}$. We then extend the notions of satisfiability, entailment and equivalence in $\mathcal{L}$ to the new language as expected (e.g., for every set $\Gamma$ of formulas of $\mathcal{L}$, $\bigvee \Gamma$ is satisfiable in an interpretation $\mathcal{M}$ if some $F \in \Gamma$ is satisfiable in $\mathcal{M}$, and so on).[6]

Let $\mathbf{F}_{\mathbf{x}}$ be the set of all formulas in the extended language above whose free variables are from the same $n$-tuple $\mathbf{x}$. One can show that mutual entailment between two formulas in $\mathbf{F}_{\mathbf{x}}$ is an equivalence relation. Let $[F]$ denote the equivalence class of a formula $F$ with respect to this relation, and let $\mathbf{E}$ denote the set of all those equivalence classes. Let $\llbracket [F] \rrbracket \overset{\text{def}}{=} \llbracket F \rrbracket$ for each $[F] \in \mathbf{E}$. The poset $\langle \mathbf{E}, \sqsubseteq_{\mathbf{E}} \rangle$ where

$$[F] \sqsubseteq_{\mathbf{E}} [G] \text{ iff } F \models_{\mathcal{L}} G$$

has a lattice structure with the following join and meet operators: $[F] \sqcup_{\mathbf{E}} [G] \overset{\text{def}}{=} [F \vee G]$ and $[F] \sqcap_{\mathbf{E}} [G] \overset{\text{def}}{=} [F \wedge G]$. It can be shown that the two functions[7]

$$\alpha_{\mathbf{E}} : \wp(\mathbf{Q}) \to \mathbf{E} \quad \overset{\text{def}}{=} \quad \lambda V. \, [\bigvee \{A_{\mathbf{v}} \mid \mathbf{v} \in V\}]$$

$$\gamma_{\mathbf{E}} : \mathbf{E} \to \wp(\mathbf{Q}) \quad \overset{\text{def}}{=} \quad \lambda E. \, \llbracket E \rrbracket$$

form a Galois connection. According to Proposition 2, the best sound abstract transformer of $f$ wrt this connection is

$$f_{\mathbf{E}} : \mathbf{E} \to \mathbf{E} \overset{\text{def}}{=} \alpha_{\mathbf{E}} \circ f \circ \gamma_{\mathbf{E}} = \lambda E. \, [\bigvee \{A_{\mathbf{v}} \mid \mathbf{v} \in \llbracket E \rrbracket\} \cup \{\mathbf{u}' \mid \mathbf{u} \in \llbracket E \rrbracket, \, \mathbf{u} \rightsquigarrow \mathbf{u}'\}\}]$$

---

[6] This is just for theoretical convenience. In practice, our method will never work with formulas $\bigvee \Gamma$ where $\Gamma$ is infinite.

[7] We borrow $\lambda$-calculus' notation to denote mathematical functions.

By our assumptions and the definition of $\alpha_{\mathbf{E}}$, the most precise abstraction of $\mathbf{I}$ is $\alpha_{\mathbf{E}}(\mathbf{I}) = [I]$.[8] It follows that in the abstract domain $\langle \mathbf{E}, \sqsubseteq_{\mathbf{E}} \rangle$ we can define the following semantics for $S$: $\mathbb{S}^{\mathbf{E}} \stackrel{\text{def}}{=} \mathrm{lfp}_{[I]}^{\sqsubseteq_{\mathbf{E}}}(f_{\mathbf{E}})$.

**Second abstraction: changing fix point computation.** For our later needs, we would like to have a fix point computation that actually enumerates the additional states discovered by the collecting semantics. The abstraction above, over-approximating sets of states by assignment formulas, is not well suited for that. Hence, we introduce another abstract transformer, on the same lattice $\langle \mathbf{E}, \sqsubseteq_{\mathbf{E}} \rangle$:

$$g_{\mathbf{E}} : \mathbf{E} \to \mathbf{E} \stackrel{\text{def}}{=} \lambda E.\, E \sqcup_{\mathbf{E}} C([A_{\nu'}] \mid T[\nu, \nu'] \text{ is sat. in } \mathcal{L},\, \nu \in [\![E]\!],\, \nu' \notin [\![E]\!])$$

where $C$ is some choice function over subsets of $\mathbf{E}$, returning one element of its input set if the set is non-empty, and $[\mathsf{false}]$ otherwise. This function maps each equivalence class $E$ to one whose extension increases $[\![E]\!]$ with just one state, chosen among the successors of the states in $[\![E]\!]$ according to the transition formula $T$. We can use $g_{\mathbf{E}}$ instead of $f_{\mathbf{E}}$ in the fix point computation thanks to the following result.

**Proposition 3 (Soundness).** *The transformers $f_{\mathbf{E}}$ and $g_{\mathbf{E}}$ have the same least-fix point greater than $[I]$, that is,* $\mathrm{lfp}_{[I]}^{\sqsubseteq_{\mathbf{E}}}(f_{\mathbf{E}}) = \mathrm{lfp}_{[I]}^{\sqsubseteq_{\mathbf{E}}}(g_{\mathbf{E}})$ .

*Proof.* Let us show that $\mathrm{lfp}_{[I]}^{\sqsubseteq_{\mathbf{E}}}(f_{\mathbf{E}}) \sqsubseteq_{\mathbf{E}} \mathrm{lfp}_{[I]}^{\sqsubseteq_{\mathbf{E}}}(g_{\mathbf{E}}) \sqsubseteq_{\mathbf{E}} \mathrm{lfp}_{[I]}^{\sqsubseteq_{\mathbf{E}}}(f_{\mathbf{E}})$. For the first, we have to show that for each element $x \in E$, we can build a $\sqsubseteq_{\mathbf{E}}$-increasing chain $X \stackrel{\text{def}}{=} x, g_{\mathbf{E}}(x), g_{\mathbf{E}}^2(x), ...$ such that its lub $\sqcup_{\mathbf{E}} X$ is greater or equal to $f_{\mathbf{E}}(x)$. Computing $f_{\mathbf{E}}(x) \wedge \neg x$ characterizes the formula describing new states, reachable in a single transition from $x$. The increasing chain $X$ is built by enumerating those elements as the one produced by the choice function $C$ in the successive application of $g_{\mathbf{E}}$.

For the second constraint, as the new element produced by the choice function $C$ is chosen among the new state reachable in a single transition, we have for all element $x \in \mathbf{E}, g_{\mathbf{E}}(x) \subseteq f_{\mathbf{E}}(x)$.                                                                    □

**Main abstraction: abstracting formulas in $\mathbf{F}_x$.** We now introduce our last abstraction, mapping formulas in $\mathbf{F}_x$ to elements of an abstract domain $\langle \mathbf{A}, \sqsubseteq_{\mathbf{A}} \rangle$ like those typically used in abstract interpretation tools (such as intervals, polyhedra, and so on).

We assume that $\mathbf{A}$ is fitted with a lattice structure with meet $\sqcap_{\mathbf{A}}$ and join $\sqcup_{\mathbf{A}}$. We also assume the existence of a *computable* monotonic concretization function $\gamma_{\mathbf{F}} : \mathbf{A} \to \mathbf{F}_x$ which associates a formula of $\mathbf{F}_x$ to each element of $\mathbf{A}$. Intuitively, we are requiring that each element of $\mathbf{A}$ be effectively representable as a formula, which in turn denotes a set of states. This requirement is easily satisfied for many numerical abstract domains and the sort of logics used in SMT. For instance, intervals can be mapped to conjunctions of inequalities between variables and values; similarly, any linear-based abstraction can be mapped to a conjunction of linear arithmetic constraints.

---

[8] Recall that $\mathbf{I}$ is the set of initial states of $S$ while $I$ is the formula denoting $\mathbf{I}$ in $\mathcal{L}$.

**Input:** $a \in \mathbf{A}$
$F[\boldsymbol{x}, \boldsymbol{x}'] := \gamma_{\mathbf{F}}(a)[\boldsymbol{x}] \wedge T[\boldsymbol{x}, \boldsymbol{x}'] \wedge \neg\gamma_{\mathbf{F}}(a)[\boldsymbol{x}']$
**if** $F$ is not satisfiable in $\mathcal{L}$ **then**
    **return** $a$
**else**
    let $\boldsymbol{v}, \boldsymbol{v}'$ be two states that satisfy $F[\boldsymbol{x}, \boldsymbol{x}']$
    **return** $a \sqcup_{\mathbf{A}} \alpha_{\mathbf{Q}}(\boldsymbol{v}')$

**Fig. 1.** Automatic abstract transformer $g_{\mathbf{A}}$.

With $\gamma : \mathbf{A} \mapsto \mathbf{E} \stackrel{\text{def}}{=} (\lambda F.\, [F]) \circ \gamma_{\mathbf{F}}$ we obtain the Galois connection

$$\alpha_\gamma : \langle \mathbf{E}, \sqsubseteq_{\mathbf{E}} \rangle \leftrightarrows \langle \mathbf{A}, \sqsubseteq_{\mathbf{A}} \rangle : \gamma$$

where, by Proposition 1, $\alpha_\gamma$ is uniquely determined by $\gamma$.

Finally, we assume the existence of a *state abstraction* function $\alpha_{\mathbf{Q}} : \mathbf{Q} \mapsto \mathbf{A}$ which directly associates states to their abstract counterparts in $\mathbf{A}$ but is such that $\alpha_\gamma([A_\nu]) \sqsubseteq_{\mathbf{A}} \alpha_{\mathbf{Q}}(\boldsymbol{v})$ for each $\boldsymbol{v} \in \mathbf{Q}$. In other words, $\alpha_\gamma$ is at least as precise as $\alpha_{\mathbf{Q}}$ when abstracting formulas satisfied by exactly one state.

**The abstract transformer.** Our main idea is to derive automatically a sound abstract transformer $g_{\mathbf{A}}$ for $g_{\mathbf{E}}$ by relying on the concretization function $\gamma$, the state abstraction $\alpha_{\mathbf{Q}}$, and a sound, complete and terminating satisfiability solver for the logic $\mathcal{L}$. We require that for each satisfiable state formula $F[\boldsymbol{x}_1, \dots, \boldsymbol{x}_k]$ the solver is able to return a state sequence $\boldsymbol{v}_1, \dots, \boldsymbol{v}_k$ satisfying $F$.

The computation of the image of an abstract element $a \in \mathbf{A}$ under $g_{\mathbf{A}}$ is described in Figure 1. Figure 2 motivates its soundness. The satisfiability tests and the choice of the states $\boldsymbol{v}$ and $\boldsymbol{v}'$ in the figure are performed by the solver for $\mathcal{L}$—which then plays for $g_{\mathbf{A}}$ the role of the choice function in the definition of $g_{\mathbf{E}}$. We point out that, while the fix point is usually computed in the abstract with the $g_{\mathbf{A}}$ function, with our approach it is not necessary to transfer back the element $a \in \mathbf{A}$ to detect the post fix point: we know that $g_{\mathbf{A}}$ has reached that point when the formula $F$ in Figure 1 is unsatisfiable.

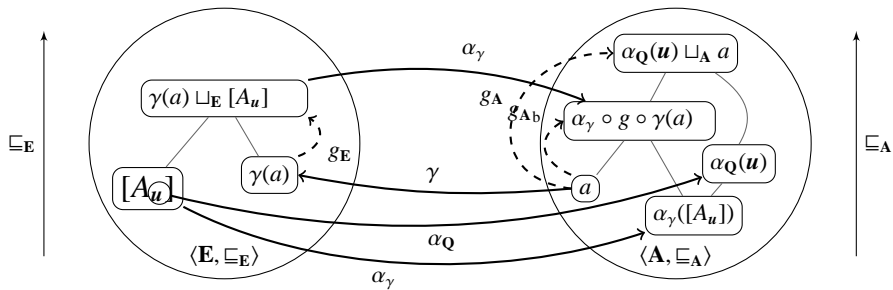

**Fig. 2.** Abstract transformer computation.

$I_\mathbf{A} := \bot$
**while** (there is a state $v$ satisfying $I[x] \wedge \neg \gamma_\mathbf{F}(I_\mathbf{A})[x]$) **do**
  $I_\mathbf{A} := I_\mathbf{A} \sqcup_\mathbf{A} \alpha_\mathbf{Q}(v)$
**return** $I_\mathbf{A}$

**Fig. 3.** Initial states over-approximation. $\bot$ is the bottom element of $\mathbf{A}$

To prove the soundness of our abstract transformer, we rely on the join-completeness property of Galois connections.

**Proposition 4 (Join-completeness for $\alpha$).** *If $\alpha : \langle A, \subseteq \rangle \leftrightarrows \langle B, \sqsubseteq \rangle : \gamma$, then $\alpha$ is join-complete, that is, $\alpha \left( \bigcup_{x \in X} x \right) = \bigsqcup_{x \in X} \alpha(x)$.*

**Theorem 1 (Soundness).** *The abstract transformer $g_\mathbf{A}$ is a sound approximation of $g_\mathbf{E}$.*

*Proof.* Figure 2 summarize the following proof elements. Let us first consider the best abstract transformer $g_{\mathbf{A}b}$ with respect to the Galois connection: for all $a \in \mathbf{A}$, $g_{\mathbf{A}b}(a) = \alpha_\gamma(g_\mathbf{E}(\gamma(a)))$: $g_{\mathbf{A}b}(a) = \alpha_\gamma(\gamma(a) \sqcup_\mathbf{E} [A_{v'}])$ where $[A_{v'}]$ is defined as the equivalence class associated with the new state $v'$ produced by the choice function. We remark that the state $v' \in \mathbf{Q}$ in the definition of $g_\mathbf{A}$ is an arbitrary new state satisfying $F$. We now have to prove that the transformer $g_\mathbf{A}$ computed by our procedure is sound with respect to the best transformer, i.e., for all $a \in \mathbf{A}$, $g_{\mathbf{A}b}(a) \sqsubseteq_\mathbf{A} g_\mathbf{A}(a)$.

Let $v' \in \mathbf{Q}$ be the new state generated by the solver. Then, $g_\mathbf{A}(a) = a \sqcup_\mathbf{A} \alpha_\mathbf{Q}(v')$. Using Property 4 on the Galois connection $(\alpha_\gamma, \gamma)$, we have that $\alpha_\gamma(\gamma(a) \sqcup_\mathbf{E} [A_u]) = \alpha_\gamma \circ \gamma(a) \sqcup_\mathbf{A} \alpha_\gamma([A_u])$. By reductivity of $\alpha_\gamma \circ \gamma$ and soundness of $\alpha_\mathbf{Q}$ with respect to $\alpha_\gamma$, we have both $\alpha_\gamma \circ \gamma(a) \sqsubseteq_\mathbf{A} a$ and $\alpha_\gamma([A_u]) \sqsubseteq \alpha_\mathbf{Q}(u)$. It follows that $\alpha_\gamma \circ \gamma(a) \sqcup_\mathbf{A} \alpha_\gamma(u) \sqsubseteq a \sqcup_\mathbf{A} \alpha_\mathbf{Q}(u)$ and $g_{\mathbf{A}b}(a) \sqsubseteq_\mathbf{A} g_\mathbf{A}(a)$.                                               □

Our eventual goal is to compute or approximate the fix point $\mathrm{lfp}_{I_\mathbf{A}}^{\sqsubseteq_\mathbf{A}}(g_\mathbf{A})$ where $I_\mathbf{A}$ is a sound over-approximation of the initial state formula or, more precisely, where $[I] \sqsubseteq_\mathbf{E} \gamma(I_\mathbf{A})$. Depending on the formula $I$ and the abstract domain $\mathbf{A}$, computing $I_\mathbf{A}$ directly from $[I]$ may not be feasible. In that case, we rely on the logic solver again to approximate $I_\mathbf{A}$. A basic algorithm for doing that is described in Figure 3. In practice, a widening operator $\nabla$ will be used in lieu of the simple join $\sqcup_\mathbf{A}$ to ensure convergence.

**Theorem 2 (Soundness).** *The element $I_\mathbf{A}$ returned by the algorithm in Figure 3 is a sound approximation of $[I]$.*

*Proof.* The initial states over-approximation $I_\mathbf{A}$ is defined as a fix point over the monotonic function adding over-approximations – through $\alpha_\mathbf{Q}$ – of new reachable states. Using Tarski's theorem, such a fix point exists. Let us consider an initial state $i \in \mathbf{Q}$ and its associate equivalence class $[A_i]$ such that is $[A_i] \not\sqsubseteq_\mathbf{E} \gamma(I_\mathbf{A})$. Then, by definition of $\gamma$ and $[\bullet]$, $i$ is not represented by the formula $\gamma_\mathbf{F}(I_\mathbf{A})$ and $I[i] \wedge \neg \gamma_\mathbf{F}(I_\mathbf{A})[i]$ is satisfiable. Then $I_\mathbf{A}$ is not a fix point.                                               □

Figure 4 summarizes our overall framework: the analysis is computed in a traditional abstract domain $\mathbf{A}$ but using a logical solver to supply a sound abstract transformer on $\mathbf{A}$.

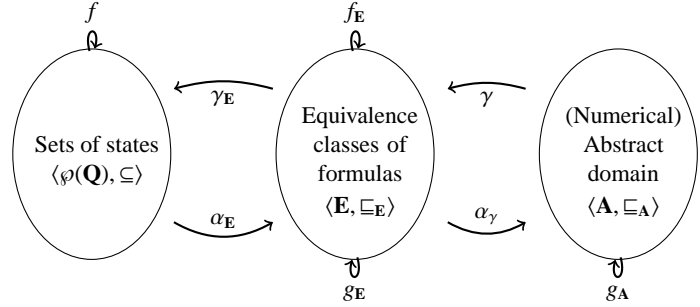

**Fig. 4.** Global framework: combination of abstractions.

## 4   On-the-fly invariant generation

A formula $F[x]$ is an *invariant for S* if $[\![F]\!]$ includes the set $R_S$ of all reachable states of $S$. Invariants have many useful applications in statical analysis, logic-based model checking, and deductive verification in general. In our abstract domain $\mathbf{E}$ from Section 3, any formula $F$ such that $\mathrm{lfp}_{[I]}^{\sqsubseteq_{\mathbf{E}}}(f_{\mathbf{E}}) \sqsubseteq_{\mathbf{E}} [F]$ is an invariant, since $R_S = [\![\mathrm{lfp}_{[I]}^{\sqsubseteq_{\mathbf{E}}}(f_{\mathbf{E}})]\!] \subseteq [\![F]\!]$.[9] By the construction of our abstraction in the domain $\mathbf{A}$, any fix point computation for the transformer $g_{\mathbf{A}} : \mathbf{A} \to \mathbf{A}$ starting with the abstract element $I_{\mathbf{A}}$ computed by the algorithm in Figure 3 produces a value $a$ such that $\gamma_{\mathbf{F}}(a)$ is an invariant for $S$.

An notable feature of our approach is that, in practice, we can modify the fix point computation for $g_{\mathbf{A}}$ to generate *intermediate* invariants, so to speak, as it goes and *before* reaching the fix point. We capitalize on the fact that $\gamma_{\mathbf{F}}(a)$ is typically a conjunction of formulas, or *properties*, $P_1, \dots, P_m$. For any intermediate value $a \in \mathbf{A}$ constructed during the fix point computation for $g_{\mathbf{A}}$, if $\gamma_{\mathbf{F}}(a) = P_1 \wedge \cdots \wedge P_m$ we can check whether any of the $P^i$'s is already invariant. This can be done, for instance, by checking that $P^i$ is inductive or, more generally, *k-inductive* [21], using the solver for $\mathcal{L}$. We discuss an efficient mechanism for doing that for multiple properties at the same time in previous work [15]. Here we point out that such a mechanism can be used in our approach to turn the abstract interpreter for $\mathbf{A}$ into an invariant stream generator.

The invariants generated in the earlier iterations of the interpreter are usually the simplest ones, e.g., bounds on a variable, and become increasingly more elaborate as the computation proceeds. The main point is that one does not need to wait until the end of a possibly complex fix point computation (using a wide sets of costly abstractions) to obtain simple invariants such as interval bounds for variables, equalities between variables and so on. Even better, the auxiliary invariants generated on the fly can be used to improve the preciseness of the very fix point computation that generated them. We can be do that by modifying the algorithm in Figure 1 to use the following strengthening

---

[9] Of course, obtaining a formula from the equivalence class $\mathrm{lfp}_{[I]}^{\sqsubseteq_{\mathbf{E}}}(f_{\mathbf{E}})$ would be enough for all purposes since that class consists of the strongest invariants for $S$. However, in general, such formulas may be infinitary or impractical to compute.

```
1  node parallel_counters (a,b,c:bool) returns(x,y: int; obs:bool);
2  var n₁, n₂:int;
3  let
4     n₁ = 10000;  n₂ = 5000;
5     x = 0 -> if (b or c) then 0 else
6             if a and (pre x) < n₁ then (pre x) + 1 else pre x;
7     y = 0 -> if c then 0 else
8             if a and (pre y) < n₂ then (pre y) + 1 else pre y;
9     obs = (x = n₁) implies (y = n₂);
10 tel
```

**Fig. 5.** Double counter example in Lustre.

of the formula $F$ defined there:

$$\gamma_{\mathbf{F}}(a)[\boldsymbol{x}] \wedge T[\boldsymbol{x}, \boldsymbol{x}'] \wedge In[\boldsymbol{x}'] \wedge \neg\gamma_{\mathbf{F}}(a)[\boldsymbol{x}'] \tag{1}$$

where, *at each call of* $g_{\mathbf{A}}$, the new subformula $In$ is the conjunction of the auxiliary invariants generated until then. This increases the precision of $g_{\mathbf{A}}$ while maintaining its soundness since it removes from consideration states that do not satisfy the current invariants (and so are necessarily unreachable).

*Example 1.* Consider the simple transition formula $T[x, x'] := (G[x] \Rightarrow x' = -1) \wedge (\neg G[x] \Rightarrow x' = x + 1)$ in a logic of integer arithmetic, for some $G$. Then suppose the current result of the fix point computation for $g_{\mathbf{A}}$ is a value $a$ with $\gamma_{\mathbf{F}}(a) = x \geq 0 \wedge P[x]$ for some sub-property $P$. Suppose also that the sub-property $x \geq 0$ has been identified as invariant. As defined in Figure 1, the computation of $g_{\mathbf{A}}(a)$ could very well produce two states $n, n'$ for $x, x'$ with $n'$ negative. Since $x \geq 0$ is invariant, both of these states are in fact unreachable. Using (1) will rule out that pair of states. □

For many of the common domains that we can use for $\mathbf{A}$, the current value $a$ of the fix point computation is actually expressed as a meet $a_1 \sqcap_{\mathbf{A}} \cdots \sqcap_{\mathbf{A}} a_k$ of other elements. Moreover, $\gamma_{\mathbf{F}}$ is meet-complete, i.e., defined so that $\gamma_{\mathbf{F}}(a_1 \sqcap_{\mathbf{A}} \cdots \sqcap_{\mathbf{A}} a_k) = \gamma_{\mathbf{F}}(a_1) \wedge \cdots \wedge \gamma_{\mathbf{F}}(a_k)$. This means that the invariant $In$ in (1) can be traced back to an $i \in \mathbf{A}$ such that $\gamma_{\mathbf{F}}(i) = In$. With this in mind, one can then understand the strengthened definition of $g_{\mathbf{A}}$ as inducing a reduced domain $\mathbf{A}'$ by the closure operator $\rho : \mathbf{A} \to \mathbf{A}' \stackrel{\text{def}}{=} \lambda a. a \sqcap_{\mathbf{A}} i$ (c.f. [7, §6]). Since the widening operators used in fix point computations are in general non-monotonic, enforcing invariants while using widening is helpful in reducing the loss of precision caused by widening.

## 5   Application: invariant generation for Lustre programs

This work was motivated by the problem of proving safety (i.e., invariant) properties of Lustre programs. Lustre [12] is a synchronous data-flow specification/programming language with infinite streams of values of three basic types: Booleans, integers, and reals. It is used to model control software in embedded devices. Properties to be proved invariant are often introduced within Lustre programs as observer Boolean streams. Checking their invariance amounts to checking that their corresponding flow is constantly true. In previous work, we have developed a $k$-induction-based parallel model checker, called Kind [17], which uses SMT solvers as its main reasoning engine. Kind benefits from the use of auxiliary invariant generators to strengthen its basic $k$-induction procedure [16]. We have implemented the fix point computation method described here as an additional on-line invariant generator for Kind.

Kind actually works with an idealized version of Lustre that treats Lustre numerical types as infinite-precision. Idealized Lustre programs can be readily recast as transition systems in a three-sorted concrete domain with Booleans, (mathematical) integers and reals. Such systems can be almost directly encoded and reasoned about in a quantifier-free logic of mixed integer and real arithmetic with uninterpreted function symbols. The linear fragment of that logic, which we could call QF_UFLIRA in the nomenclature of SMT-LIB [1], can be efficiently decided by most major SMT solvers.[10]

This means that Lustre programs limited to linear arithmetic are amenable to analysis with our method. As abstract domain we use one defined, as usual, as a reduced product of a varieties of abstract domains, including relational and non-relational ones—partitioning mechanisms allow our tool to express some non-linear properties. Our implementation of the function $\gamma_{\mathbf{F}}$ converts abstract elements into formulas of QF_UFLIRA as one would expect: an interval $[a;b]$ for a variable $x$ is converted into the formula $a \leq x \wedge x \leq b$; a linear constraint $\Sigma_i\, a_i \cdot x_i \geq c$ is mapped directly to the corresponding formula of QF_UFLIRA. The translation is extended homomorphically to more complex elements. For instance, elements that are the meet of other ones (such as polyhedra, etc.) are converted to the conjunction of the conversion of the components.

Our implementation is written in OCaml, relies on the APRON abstract domain library [14], and shares with Kind, also written OCaml, modules to encode Lustre programs as transition systems in the QF_UFLIRA logic and to interact with the SMT solver.

*Example 2.* Let us illustrate the use of our invariant generator on a typical example: counters, which are use widely within safety mechanisms for critical systems.[11] In the Lustre program shown in Figure 5, two counters $x$ and $y$ are incremented up to their respective maximum value whenever the input value $a$ is **true**; both are reset to 0 when the input $c$ is **true**. The counter $x$ is reset also when the input $b$ is **true**. Suppose we would like to prove that whenever $x$ reaches its maximum value, so does $y$. This property is expressed by the synchronous observer *obs*. It is enough to show then that the Boolean stream *obs* is equal to the constant stream **true**.

The invariant generator discovers without any special tuning the fact that $x \in [0; 10000]$ and $y \in [0; 5000]$. With additional partitioning parameters, it can in fact generate the target property itself: $x = n_1 \Rightarrow y = n_2$. Focusing on $x$ and $y$, the only two stateful variables in the program,[12] Figure 6 shows the first four states enumerated

---

[10] That includes CVC3 and Yices [2,8], the SMT solvers used by Kind.

[11] The example and the tools described here can be found at `http://clc.cs.uiowa.edu/Kind/SAS12` .

[12] The Lustre expression (`pre` $x$) denotes the value of $x$ in the preceding state.

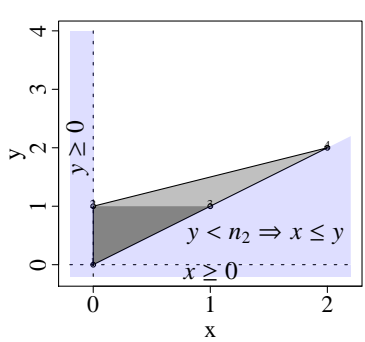

The injected states are, in order, $(0, 0)$, $(0, 1)$, $(1, 1)$ and $(2, 2)$. After the injection of $(1,1)$, the current abstract element is described by the dark triangle: $0 \leq x \leq 1$, $0 \leq y \leq 1$ and $x \leq y$. When using our partitioned analysis (with explicit partitioning), we also obtain properties under the implication: $y < n_2 \Rightarrow \ldots$ At the fourth iteration, three sub-properties of the one expressed by the dark triangle are proven invariant: $0 \leq x$, $0 \leq y$ and $y < n_2 \Rightarrow x \leq y$. Those invariants immediately communicated to Kind's $k$-induction engine, but also used internally to constrain the following iterations.

**Fig. 6.** First four steps of the fix point computation for the example.

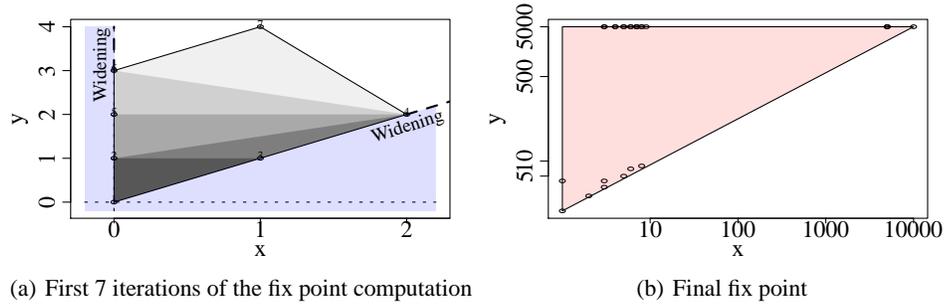

(a) First 7 iterations of the fix point computation

(b) Final fix point

**Fig. 7.** Iterations and final fix point for example.

by our fix point algorithm and injected into the abstract domain. At the forth iteration of the computation the following properties are already identified as invariant: $x \geq 0$, $y \geq 0$ and $y < n_2 \Rightarrow x \leq y$.

Figure 7 shows another intermediate element obtained before widening, as well as the final abstract element obtained, the fix point of the abstract collecting semantics. On this example, using $k$-induction alone Kind is not able to prove in reasonable time the property expressed by *obs*. It principle it could, but since the property is 10000-inductive, $k$-induction requires too many unrollings of the system's transition relation to scale. However, using the auxiliary invariant $y < n_2 \implies x \leq y$ produced at the forth iteration of our invariant generator, along with the bounds obtained without any partitioning, Kind is able to prove the target property instantaneously.

## 6   Conclusion and further work

The framework we presented offers two main contributions: (i) a systematic and automatic generation of abstract transformers relying on logic solvers and abstract domain libraries; (ii) the generation of invariants during the computation of post fix points. Although this paper focused mainly on least fix point computations of a forward semantics, the approach can be applied in a wide range of settings: computing a greatest fix point or analyzing a backward semantics is directly expressible in the framework without major modifications. This approach is truly automatic whenever the target system can be encoded in a suitable decidable logic, and abstract domain elements are representable in that logic. Such conditions are often easy to satisfy for systems already analyzable with SMT solvers, and for numerous available abstract domains. Under those conditions one obtains an abstract interpreter for free. There are no restrictions on the system's language constructions handled or on the specific abstract domains that could be used. Furthermore, our framework facilitates the expression of big step semantics (on the logical side) and therefore avoids the loss of precision obtained when applying abstract transfer functions at a small step semantics level.

To our knowledge, our initial implementation of the framework is the only available tool based on abstract interpretation and Kleene-style fix point computation that provides invariants before the post fix point is reached. In a multi-analyzer setting, the possibility to share invariants before the end of the computation can drastically increase performance. But that sort of intermediate but guaranteed information could be extremely valuable even in a standalone use. For example, when analyzing a 200k-loc critical embedded software for the absence of run time errors [6], one could observe during the computation the sections of the code that are already proven (e.g., no division by zero at a certain statement) or (false) alarms. This contrasts with the current general practice where one has to wait, possibly for hours, for the fix point computation to end before interpreting the results, and seeing perhaps that certain parameters need further tuning.

We have implemented our method and applied it to Lustre programs in the context of a larger project on the analysis of synchronous systems. Further work will involve a more extensive experimental evaluation of the method to assess its benefits on a larger scale.